\documentclass[showpacs,preprintnumbers]{revtex4}
\usepackage{amsfonts}
\usepackage{amssymb}
\usepackage{graphicx}
\usepackage{amsmath,float}
\usepackage{epsfig}
\usepackage{array}
\usepackage{color}	
\usepackage[T1]{fontenc}
\allowdisplaybreaks
\begin{document}
	
\title{Charged Strange Star Model with Stringy Quark Matter in Rainbow Gravity}

\author{Wasib Ali}
\email{msf2100636@ue.edu.pk}
\affiliation{Department of Mathematics, Division of Science and Technology, University of Education, Lahore, Pakistan.}

\author{Umber Sheikh}
\email{umber.sheikh@ue.edu.pk}
\affiliation{Department of Mathematics, Division of Science and Technology, University of Education, Lahore, Pakistan.}


\allowdisplaybreaks

\begin{abstract}
This study deals with the formation and evolution of a strange star in the Krori Barua Rainbow spacetime from collapsing charged stringy quark matter. The dynamical variables are explored from the field equations, taking into account the effects of particle's energy on the mass density, pressure, and string tension. The electric field is also computed using the MIT Bag model. The real time data of $SAX~J1808.4-3658$ is used to analyzed the physical properties including gradients, energy conditions, anisotropy, stability, Tolman Oppenheimer Volkoff equation, mass function, compactness, and red-shift. The graphical analysis has been made according to both the theories of rainbow gravity and general relativity. The energy conditions and anisotropy are found to be satisfied, indicating the physical existence of suggested model. Tolman Oppenheimer Volkoff equation is satisfied indicating equilibrium of forces and stability of the compact object. Overall, our model is consistent with the observational information of $SAX~J1808.4-3658.$ 
\end{abstract}

\maketitle


\section{Introduction}

Compact stars are astrophysical objects that captivate researchers due to their intriguing properties and their potential to shed light on the fundamental physics of matter and gravity. These celestial objects, characterized by their high mass and compactness, possess immense gravitational fields that challenge our current understanding of the laws governing the universe. They provide a unique laboratory for studying matter under extreme conditions and exploring the nature of gravity itself.

Among the various types of compact stars, strange stars have attracted significant attention. Strange stars are objects composed primarily of up, down, and strange quark matter. These exotic objects are believed to exist in the universe and have been proposed as an alternative to neutron stars or black holes. The existence of uncharged and charged strange stars is analyzed in the background of torsion based $f(T)$ modified gravity with the discussion on maximum mass limit \cite{1,2}.

The physical properties and existence of charged anisotropic strange stars are investigated within the Rastall framework \cite{3}. The presence of strange stars within the $f(T, T)$ gravity reveals the properties under the influence of torsion \cite{4}. A unique anisotropic solution for the strange star model in the context of 5D Einstein-Gauss-Bonnet gravity is examined considering the implications of higher-dimensional gravitational theories \cite{5}. Solving the field equations within the context of $f(R, T)$ gravity and the discoveries contributed significantly to the theoretical understanding of strange stars \cite{6}. A comprehensive analysis is conducted of various candidates for strange stars, including $EXO 1785-248, LMC X-4, SMC X-1, SAX J1808.4-3658, 4U 1538-52,$ and $Her X-1$. The physical features of these candidates, considering them as uncharged and charged, static, spherically symmetric, and isotropic fluid spheres, thus valuable insights are provided into their unique characteristics \cite{7,c1,c2,c3,c4}. 

Krori Barua (KB) metric is exemplary to present a compact star as this metric does not contain any singularity. It is found to be feasible to present the famous neutron, strange, and quark stars using this metric in different theories of gravity. The physical properties and stability of anisotropic compact stellar objects in energy-momentum squared gravity using KB metric are verified for $Her X-1,~4U 1538-52,$ and $SAX J1808.4-3658$ compact stars \cite{c10}. The values of the Rastall parameter for $SAXJ1808.4-3658,~VelaX-12,$ and $HerX-1$ are investigated for their physical existence using KB metric \cite{c11}. The real-time data for $PSRJ1614-2230,~4U1608-52,~CenX-3,~EXO1785-248,$ and $SMCX-1$ has been verified using KB metric in $f(T, T)$ theory \cite{c12}. The physical properties of $Her-1,~SAXJ1808.4-3658,~4U1820-30,~PSRJ1614-2230,~VelaX-1,$ and $CenX-3$ are verified using KB metric with Hu-Sawicki and logarithmic model in $f(G, R)$ gravity  \cite{c13}. The same has been done for generic data in higher-order torsion theory \cite{c14}.

Rainbow gravity (RG) is a framework that proposes modifications to general relativity by amalgamating energy-dependent spacetime curvature. The notion of rainbow gravity stems from the concept that the effects of quantum gravity may manifest at energy scales close to the Planck scale. A ubiquitous conformal coupling results from the effective switch-off of RG at higher or at the very least energies. This suggests a significant departure from the standard behavior of gravity in the high-energy regime \cite{9}. The hydrostatic equilibrium and structural properties of white dwarfs in Rastall-Rainbow gravity are investigated and aimed to understand the effects of this modified gravitational theory \cite{10}. The thermodynamics and phase transition phenomena associated with rainbow Schwarzschild black holes were examined utilizing the rainbow functions, shedding light on the modified black hole properties \cite{11}.  The thermodynamic properties and behavior of black holes are discussed \cite{13}. The dynamics and properties of isotropic quantum cosmological perfect fluid provide insights into the modified gravitational effects at the quantum cosmological level \cite{12}.

RG theory is an effort to merge general relativity and quantum mechanics. When we deal with strange matter \cite{c21}, the radiation emerging from the black hole \cite{c22,c24,c25}, and information paradox like problems \cite{F1}-\cite{F4}. The equilibrium configurations of neutron stars in RG demonstrated that the energy-dependent curvature can significantly affect the mass-radius relationship \cite{A1}. The stability of strange quark stars in RG, concluding that the modified curvature leads to a more stable configuration \cite{A2}. These studies highlight the relevance and potential of RG in understanding the behavior of compact stars.

In this manuscript, we will examine the effect of probing particle's energy, which, when approaching zero, yields results consistent with general relativity. The paper is structured as follows: The next section provides a detailed explanation of the KB Rainbow geometry, focusing on its application to the star and string fluid matter distribution. Section 3 will present the formulation of the field equations, along with the solutions for the dynamical variables and the analysis of energy conditions. In Section 5, we will model the physical characteristics of strange stars, utilizing the specific physical data of $SAXJ1808.4-3658$. Lastly, the concluding section will summarize our findings and provide insights into the overall study.

\section{Geometry of Spacetime and Fluid Description}

We have assumed a star filled with charged stringy quark matter. This matter obeys the energy-momentum tensor \cite{F3}:
\begin{equation}\label{sf}
	T_{\mu\nu}=(\rho+p)u_{\mu}u_{\nu}-pg_{\mu\nu}-l x_{\mu}x_{\nu} + \mathbb{E}^2 g_{\mu\nu},
\end{equation}
where $\rho,~p,~l,~E^2$ represents the matter density, fluid pressure, string tension, and squared electric field intensity respectively. The electric field intensity can be described in terms of charge $Q$ as $\mathbb{E}^2=\frac{Q^2}{4\pi r}.$
Position and velocity unit vectors are presented by $x_{\mu}$ and $u_{\mu}$ respectively. It is noteworthy that for Eq.(\ref{sf}):
\begin{itemize}
	\item $p=0$ gives a string cloud, 
	\item $l=0$ leads to quark matter,
	\item $\rho-p=0$ provides stringy stiff fluid,
	\item $\rho+p=0$ exhibits stringy anti-stiff fluid, 
	\item $\rho+l=0$ presents a Reddy strings oriented fluid,
	\item $\rho-l=0$ shows a Nambu strings oriented fluid respectively. 
\end{itemize}
Stringy quark fluid is expected to be present in early Universe \cite{PKS}. The stringy quarks are bound under pressure and form the stringy quark fluid. We have assumed radially stretched one-dimensional strings attached with each quark in the fluid.  

This matter is assumed to collapse in the interior of a star with the Karori Barua rainbow (KBR) metric:
\begin{equation}\label{IS}
	ds_-^2=-\frac{1}{s^2\left(\frac{E}{E_P}\right)}e^{AR^2+C}dt^2+\frac{1}{m^2\left(\frac{E}{E_P}\right)}\left(e^{YR^2}dr^2
	+r^2d\Omega^2\right).
\end{equation}
with $d\Omega^2=d\theta^2+\sin^2\theta d\phi^2.$ As the interior spacetime is filled with charged fluid, this leads to a charge exterior spacetime for matching at the surface of the star. Thus, we have assumed the exterior as Reissner Nordstrom line element expressed as:
\begin{equation}\label{ES}
	ds_+^2=-\frac{1}{s^2\left(\frac{E}{E_P}\right)}(1-2\frac{M}{r}+\frac{Q^2}{r^2})dt^2+\frac{1}{m^2\left(\frac{E}{E_P}\right)}\left((1-2\frac{M}{r}+\frac{Q^2}{r^2})^{-1}dr^2+r^2d\Omega^2\right),
\end{equation}
where $M$ and $Q$ demonstrate the configuration's mass and charge.

It is worth noting that in both the line elements, the functions $s\left(\frac{E}{E_P}\right)=1$ and $m\left(\frac{E}{E_P}\right)=\sqrt{1-\eta\frac{E}{E_P}}$ are the rainbow functions. Here the rainbow parameter $\eta>0,$ $E,~E_P$ are the probing particle's and Planck's energies. The pair of rainbow functions can be used to illustrate the dynamics of black holes and the information paradox. Furthermore, the quantum radiations produced by gamma-ray bursts are anticipated to reveal detailed information on the strange star's behavior.

Comparing Eqs. (\ref{IS}) and (\ref{ES}), we reached at the following relations at the boundary of the strange star:
\begin{equation}
	1-2\frac{M}{R}+\frac{Q^2}{R^2}=e^{AR^2+C},~(1-2\frac{M}{R}+\frac{Q^2}{R^2})^{-1}=e^{YR^2},~\frac{M}{R^2}-\frac{Q^2}{R^3}=ARe^{AR^2+C}, 
\end{equation}
leading to
\begin{equation}
	Y=-\frac{1}{R^2}\ln\left(1-2\frac{M}{R}+\frac{Q^2}{R^2}\right),~
	A=\frac{\left(\frac{M}{R}-\frac{Q^2}{R^2}\right)}{R^{2}\left(1-2\frac{M}{R}+\frac{Q^2}{R^2}\right)},~
	C=\ln\left(1-2\frac{M}{R}+\frac{Q^2}{R^2}\right)-\frac{\left(\frac{M}{R}-\frac{Q^2}{R^2}\right)}{\left(1-2\frac{M}{R}+\frac{Q^2}{R^2}\right)}.
\end{equation}

\section{Formulation of Field Equations and Dynamical Variables}

The famous Einstein field equations of general relativity relate astronomical objects' geometry to their physics. Because RG incorporates all of the essential ideas of general relativity, with minor departures from quantum physics, Einstein's field equations are valid in this theory to describe collapse.

The Einstein field equations for collapsing stringy quark fluid (Eq.(\ref{sf})) in KBR spacetime (Eq.(\ref{IS})) are computed in geometric units as:
\begin{eqnarray}\label{EFE1}
&&G_{tt}:\frac{-m^2 e^{Ar^2+C}(2Yr^2+e^{Yr^2}-1)}{s^2r^2e^{Yr^2}}=-\frac{8\pi e^{Ar^2+C}}{s^2}(\rho+\mathbb{E}^2),\\
\label{EFE2}
&&G_{rr}:\frac{-2Ar^2+e^{Yr^2}-1}{r^2}=\frac{-8\pi e^{Yr^2}}{m^2}(l+p-\mathbb{E}^2),\\
\label{EFE3}
&&G_{\theta\theta}:\frac{r^2(YAr^2-A^2r^2+Y-2A)}{e^{Yr^2}}= \frac{8\pi r^2}{m^2}(\mathbb{E}^2-p),\\
\label{EFE4}
&&G_{\phi\phi}:r^2\sin^2\theta\frac{YQr^2-Q^2r^2+Y-2Q}{e^{Yr^2}}= \frac{8\pi r^2 \sin^2\theta}{m^2}(\mathbb{E}^2-p).
\end{eqnarray}
It is noteworthy that Eqs.(\ref{EFE3}) and (\ref{EFE4}) are found to be the same. 

The famous MIT bag model suggests the quark matter density and fluid's pressure admits the relation: $p_r=\frac{1}{3}(\rho-4B_g),p_r=p+l.$ Using this model, 
Eqs.(\ref{EFE1}),(\ref{EFE2}) and (\ref{EFE3}) respectively lead to the density $\rho$, pressures $p,$ string tension $l,$ and squared electric field intensity $E^2$ of the stringy quark matter is computed as:
\begin{eqnarray}\label{D}
\rho&=&\frac{3m^2}{16\pi e^{Yr^2}}(A+Y)+B_g,\\\label{P}
p&=&\frac{m^2}{8\pi e^{Yr^2}}\left(\frac{e^{Yr^2}-1}{r^2}+\frac{1}{2}(A-Y)+ (A-Y)Ar^2\right)-B_g,\\\label{ST}
l&=&\frac{m^2}{8\pi e^{Yr^2}}(YAr^2-A^2r^2+Y+\frac{1-e^{Yr^2}}{r^2}),\\\label{E}
\mathbb{E}^2&=&\frac{m^2}{8\pi e^{Yr^2}}\left(\frac{e^{Yr^2}-1}{r^2}+\frac{1}{2}(Y-3A)\right)-B_g.
\end{eqnarray}
It is visible that all these dynamical variables are radius dependent quantities. The rainbow function $m=m\left(\frac{E}{E_P}\right)$ depends on the probing particle's energy $E.$ The MIT bag constant is $B_g=2.543\times 10^{-4}$. The literature reveals that $SAXJ1808.4-3658$ admits mass $M=1.435\bigodot$ where $\bigodot$ presents sun's mass, and radius $R=7.07km$ \cite{22b}. This leads to the calculation of the KB constants as $A=1.282\times10^{-2}$ and $Y=1.702\times10^{-1}.$

The density $\rho$, pressures $p,$ string tension $l$ and squared electric field intensity $E^2$ of the stringy quark matter are exhibited graphically for the region $0\leq R\leq 5,~0\leq E\leq 5.$
\begin{figure}[H]
\centering
\includegraphics[width=0.5\linewidth]{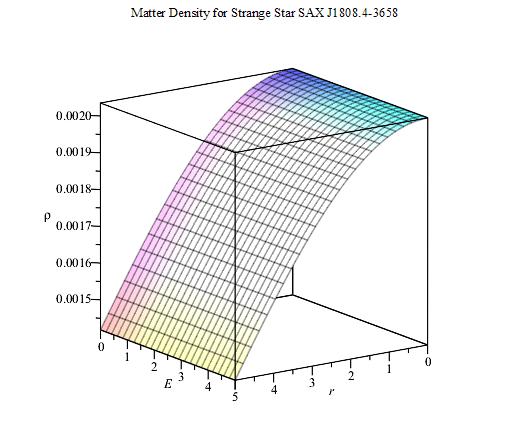}
\caption{Progression of mass density in RG for $SAX.J1808.4-3658$.}
\label{f1}
\end{figure}
FIG. \ref{f1} depicts the mass density of a collapsing fluid. The mass density is highest and finite in the star's center. However, as one moves radially out from the center, mass density drops. Furthermore, particles with more energy have a higher mass density than particles with less energy

\begin{figure}[H]
\centering
\includegraphics[width=0.5\linewidth]{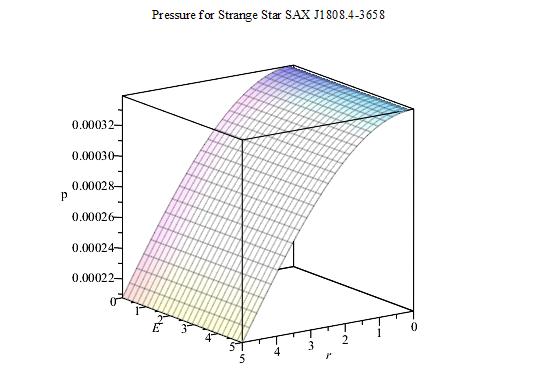}
\caption{Progression of pressure in RG for $SAX.J1808.4-3658$.}
\label{f2}
\end{figure}

The pressure behavior of a collapsing fluid is seen in FIG. \ref{f2}. The graph shows that the pressure is limited and is greatest close to the star's core. When it gets close to the star's outer edge, it descends. As the energy of the probing particle increases, pressure increases.

\begin{figure}[H]
	\centering
	\includegraphics[width=0.5\linewidth]{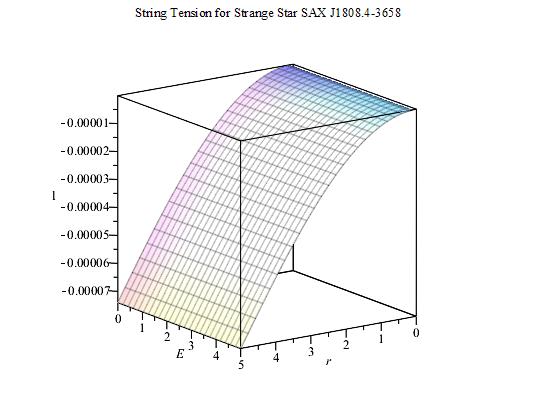}
	\caption{Progression of string tension in RG for $SAX.J1808.4-3658$.}
	\label{f3}
\end{figure}
A collapsing fluid's string tension is seen in FIG. \ref{f3}. String tension is negative and increases with decreasing radius of the star. As the energy of the probing particle rises, string tension rises as well.

\begin{figure}[H]
	\centering
	\includegraphics[width=0.55\linewidth]{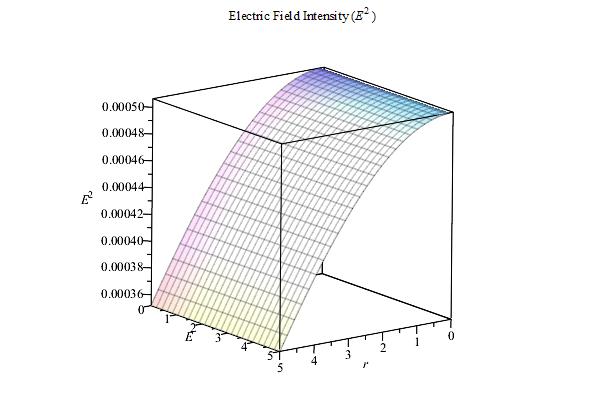}
	\caption{Progression of string tension in RG for $SAX.J1808.4-3658$.}
	\label{f4}
\end{figure}
FIG. \ref{f4} exhibits the squared electric field intensity behavior of a collapsing fluid. The graph indicates that electric field intensity is finite and highest near the star's center. It decreases when it approaches the star's outer boundary.

In general relativity, we may also examine the behavior of these variables by calculating the energy of a vanishing probing particle. The following is general relativistic interpretation of these quantities:
\begin{figure}[H]
	\centering
	\includegraphics[width=0.33\linewidth]{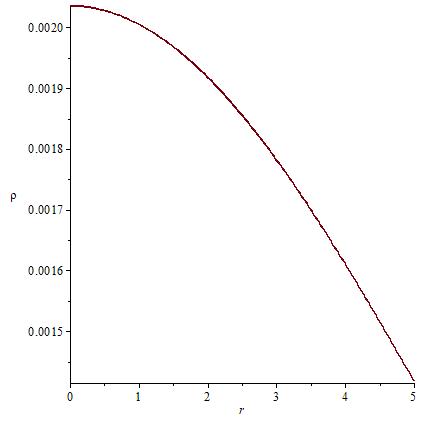}
	\includegraphics[width=0.33\linewidth]{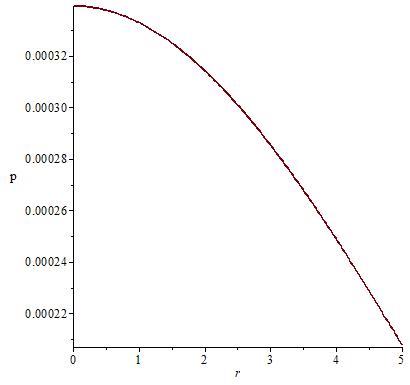}
\end{figure}
\begin{figure}[H]
	\centering
		\includegraphics[width=0.33\linewidth]{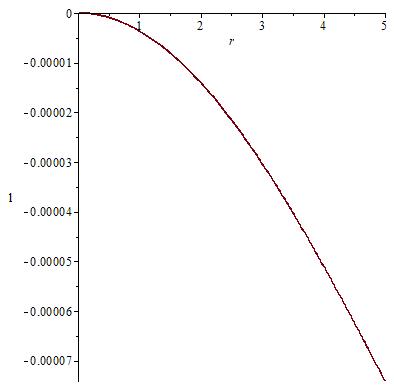}
		\includegraphics[width=0.33\linewidth]{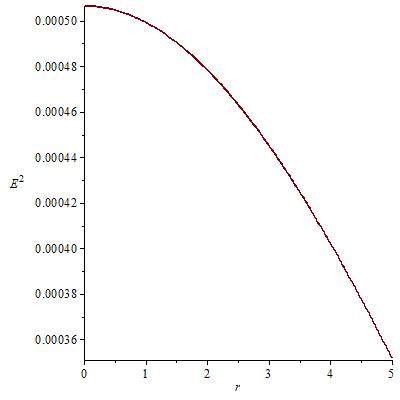}
	\caption{Progression of matter density, pressure and string tension in GR for $SAX.J1808.4-3658$.}
	\label{f5}
\end{figure}
FIG. \ref{f5} exhibits matter density, pressure, and string tension evolution. It is obvious that all of the quantities except electric field intensity reach their highest levels in the distribution center. However, when the radius rises, these variables diminish, where as the electric field intensity increases with the increase of radius. This results in the formation of a compact star as a result of collapsing string fluid.

The distribution of mass at the star's periphery and the massive concentration at its center are seen in FIGS. \ref{f1}-\ref{f4}. The creation of a compact star is indicated. The fact that the center of resulting astronomical object has a finite maximum output of mass density, pressure, and string tension indicates that it is a compact star rather than a black hole.

\subsection{Gradients}

For the sake of a viable strange star, the dynamical variables are required to possess the following gradients:
\begin{itemize}
	\item $\frac{d\rho}{dr}\leq0,$
	\item $\frac{dp}{dr}\leq0,$
	\item $\frac{dl}{dr}\leq0.$
\end{itemize}

By calculating the derivative of the three dynamical quantities $\rho,~p,~l$ (Eqs.(\ref{D})-(\ref{ST})) corresponding to radius, we can derive the following gradients $\frac{d\rho}{dr},~\frac{dp}{dr},~\frac{dl}{dr}$ as:
\begin{eqnarray}\label{dr}
\frac{d\rho}{dr}&=&-\frac{3Yrm^2\left(Y+A\right)}{8{\pi}\mathrm{e}^{Yr^2}},\nonumber\\\label{dp}
\frac{dp}{dr}&=&-\frac{m^2\left(2\mathrm{e}^{Yr^2}+\left(-Y^2+3AY-2A^2\right)r^4+\left(2A^2Y-2AY^2\right)r^6-2Yr^2-2\right)}{8{\pi}r^3\mathrm{e}^{Yr^2}},	\nonumber\\\label{dl}
\frac{dl}{dr}&=&-\frac{m^2}{4\pi r^3e^{Yr^2}}(1+Yr^2-e^{Yr^2}+(A^2-YA+Y^2)r^4-YA(A-Y)r^6).\nonumber
\end{eqnarray}

The general relativistic analogue of these equations are:
\begin{eqnarray}
\frac{d\rho}{dr}|_{E=0}&=&-\frac{3Yr\left(Y+A\right)}{8{\pi}\mathrm{e}^{Yr^2}},\nonumber\\
\frac{dp}{dr}|_{E=0}&=&-\frac{\left(2\mathrm{e}^{Yr^2}+\left(-Y^2+3AY-2A^2\right)r^4+\left(2A^2Y-2AY^2\right)r^6-2Yr^2-2\right)}{8{\pi}r^3\mathrm{e}^{Yr^2}},\nonumber\\
\frac{dl}{dr}|_{E=0}&=&-\frac{\left(1+Yr^2-e^{Yr^2}+(A^2-YA+Y^2)r^4-YA(A-Y)r^6\right)}{4\pi r^3e^{Yr^2}}.\nonumber\\
\end{eqnarray}

The first order derivatives of the dynamical quantities $\rho,~p,~l$ with respect to the radius are shown as follows:
\begin{figure}[H]\centering
\includegraphics[width=0.33\linewidth]{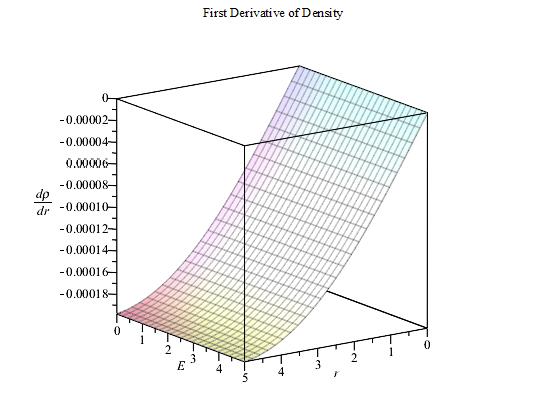}
\includegraphics[width=0.33\linewidth]{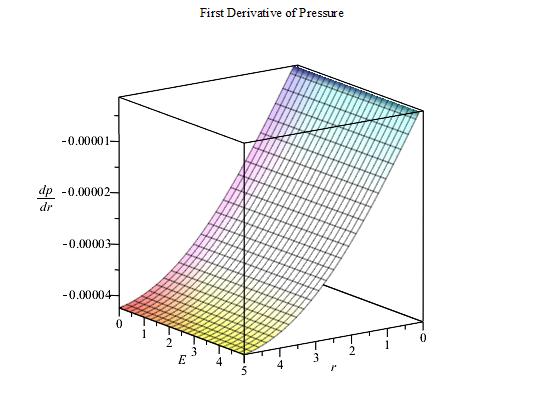}
\includegraphics[width=0.32\linewidth]{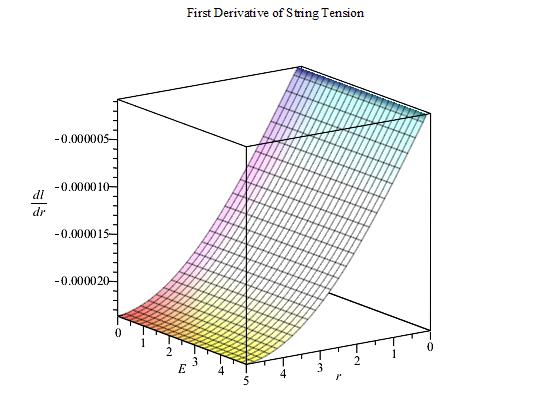}
\caption{Progression of $\frac{d\rho}{dr},~\frac{dp}{dr},~\frac{dl}{dr}$ in RG for $SAX.J1808.4-3658$.}
\label{f6}
\end{figure}

FIG. \ref{f6} shows the progression of $\frac{d\rho}{dr},~\frac{dp}{dr},~\frac{dl}{dr}$ corresponding to the radius and probing particle's energy. The graph clearly shows that this change is greatest near the center. As $r$ increases, this change decreases exponentially. Furthermore, when $E$ grows, the gradients of dynamical quantities reduce. As a result, a particle with higher energy sees a bigger change in dynamical quantities associated with $r$ than a particle with less energy.

\begin{figure}[H]\centering
\includegraphics[width=0.25\linewidth]{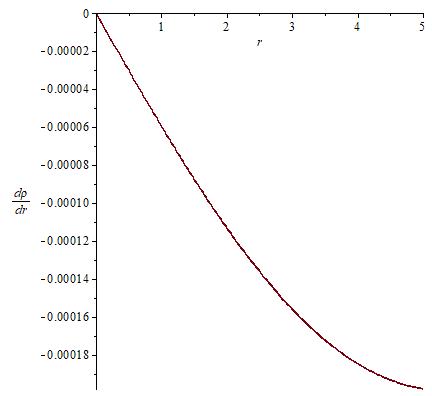}
\includegraphics[width=0.25\linewidth]{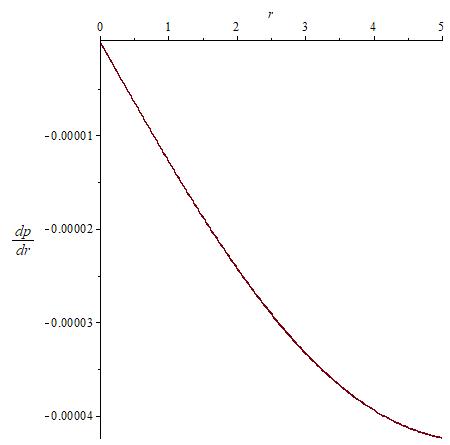}
\includegraphics[width=0.225\linewidth]{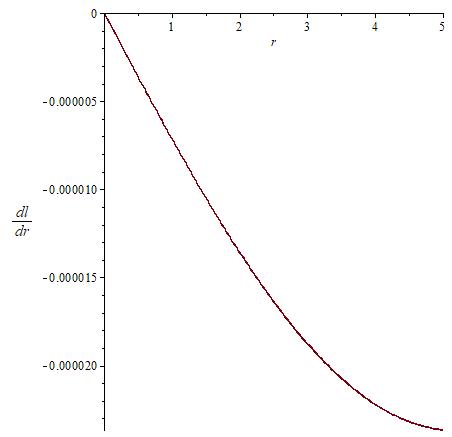}
\caption{Progression of $\frac{d\rho}{dr},~\frac{dp}{dr},~\frac{dl}{dr}$ in GR for $SAX.J1808.4-3658$.}
\label{f7}
\end{figure}

FIG. \ref{f7} depicts the progression of $\frac{d\rho}{dr},~\frac{dp}{dr},~\frac{dl}{dr}$ in GR. The graph indicates that the change is greatest in the center and lessens as we approach the star's outer border. 

\subsection{Energy Conditions}

The imposition of energy conditions identifies the physical significance of matter presented by the stress-energy tensor. Null, weak and string energy conditions  for the string fluid are $\rho+p-l>0,~\rho>0,$ and $\rho+3p-l>0$ respectively \cite{U}. These can be computed for $SAX~J1808.4-3658$ taking account of the three computed dynamical quantities, mass density $\rho,$ fluid's pressure $p,$ and string tension $l$ from Eqs.(\ref{D})-(\ref{ST}) as:

\begin{figure}[H]
	\centering
	\includegraphics[width=0.35\linewidth]{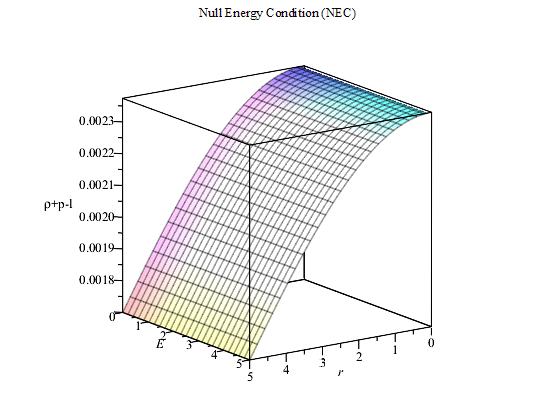}
	\includegraphics[width=0.3\linewidth]{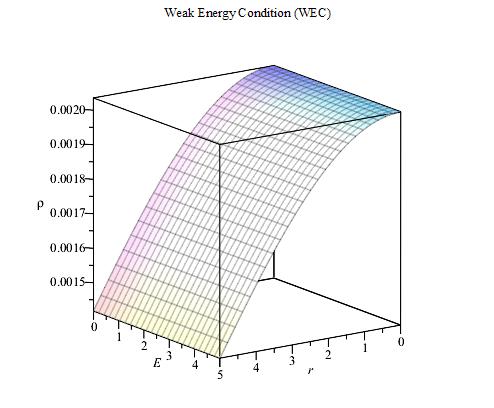}
	\includegraphics[width=0.35\linewidth]{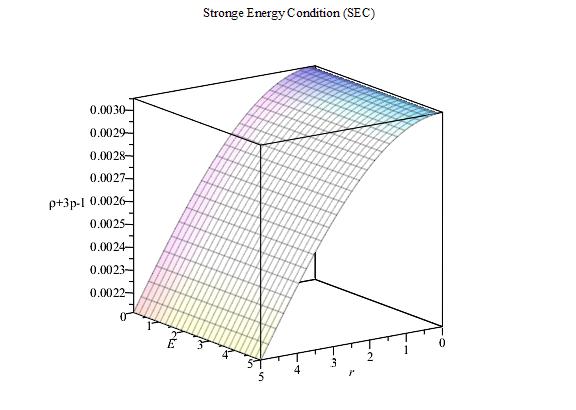}
	\caption{Progression of null, weak and strong energy expressions in RG for $SAX.J1808.4-3658$.}
	\label{f8}
\end{figure}

FIG. \ref{f8} shows that mass density, pressure and the string tension we derived for $SAX~J1808.4-3658$ satisfy the null, weak, and strong energy conditions. This means that matter is non-negative when it encounters a time-like or light-like vector. Furthermore, matters is drawn towards the gravity. These characteristics validate the physical reality of our stringy quark matter model.

\begin{figure}[H]
	\centering
	\includegraphics[width=0.225\linewidth]{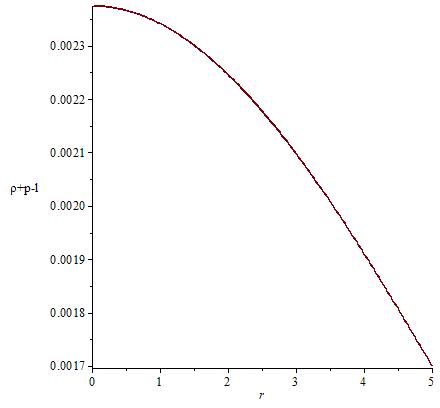}
	\includegraphics[width=0.25\linewidth]{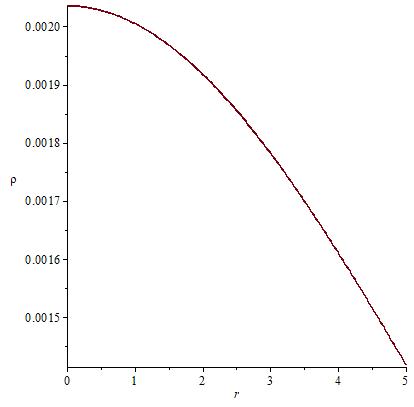}
	\includegraphics[width=0.22\linewidth]{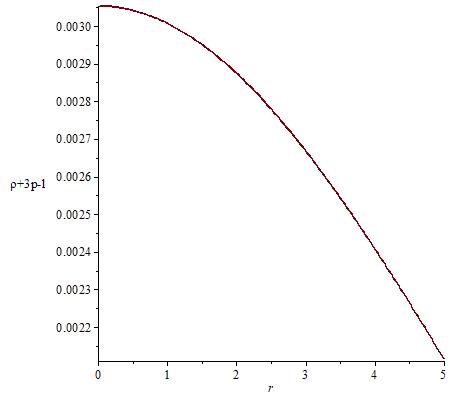}
	\caption{Progression of null, weak and strong energy expressions in GR for $SAX.J1808.4-3658$.}
	\label{f18}
\end{figure}

All of the energy conditions have been satisfied in GR, as shown in FIG. \ref{f18}, confirming the presence of string fluid in GR.

\section{Physical Properties}

Anisotropy, EOS parameters, the TOV equation, Harrera's cracking condition, mass function, compactness, and redshift are just a few of the physical characteristics of $SAX.J1808.4-3658$ that will be covered.

\subsection{Anisotropy}

Anisotropy is the property of compact star models which assume various characteristics in different directions. It is the essential component of the accurate compact star model. The existence of distinct fluid characteristics, such as a magnetic, external field, rotation, viscosity, and phase transitions, can cause anisotropy in a fluid's pressure. In a collapsing process, anisotropy is the force necessary to keep the compact star from falling farther. It should be noted that raising anisotropy parameters improves zone stability. Using Eqs. (\ref{P}) and (\ref{ST}) anisotropy of the spherical distribution of matter in KBR metric is  determined to be:
\begin{equation}\label{anisorg}
	\Delta=p_t-p_r=p-(p+l)=-\frac{m^2}{\kappa e^{Yr^2}}\left(YAr^2-A^2r^2+Y-\frac{e^{Yr^2}+1}{r^2}\right).
\end{equation}
Anisotropy has the same general relativistic equivalent as described above.

Following is a representation of $SAX.J1808.4-3658$'s anisotropy:
\begin{figure}[H]
	\centering	
	\includegraphics[width=0.5\linewidth]{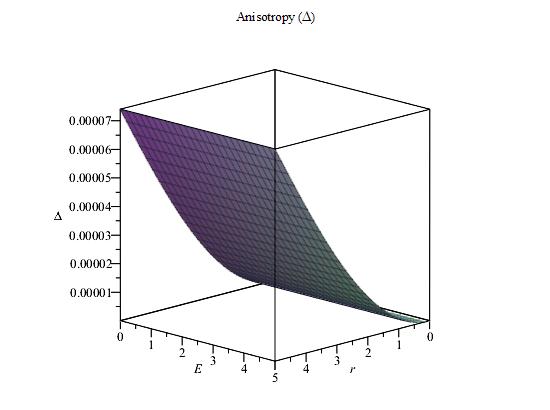}
	\includegraphics[width=0.35\linewidth]{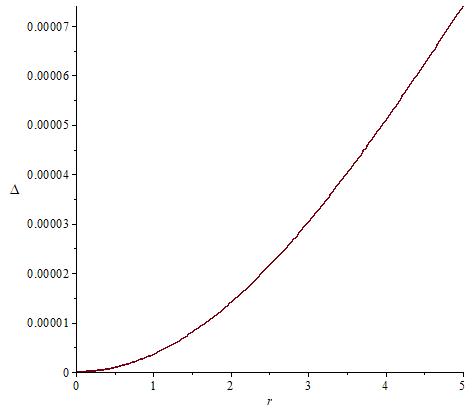}
	\caption{Progression of anisotropy in RG and GR for $SAX.J1808.4-3658$.}
	\label{f9}
\end{figure}

FIG. \ref{f9} shows that at the center of the star, the anisotropy is decreasing. It steadily grows as we approach the star's outskirts. An analogue of anisotropy in GR is obtained by replacing $E=0$, which yields the same result as in RG.

\subsection{Equation of State Parameters}

The EOS describes the state of fluids under certain conditions. An efficient knowledge of the EOS parameters helps to understand the processes in extreme conditions. There are two EOS parameters for string fluid:
\begin{eqnarray}
	\textmd{Radial~Parameter:} w_r&=&\frac{m^2(A+Y)-16\pi e^{Yr^2}B_g}{3m^2(A+Y) +16\pi e^{Yr^2}B_g},\nonumber\\
	\textmd{Tangential~Parameter:} w_t&=&\frac{2m^2\left(\frac{e^{Yr^2}-1}{r^2}+ (A-Y)Ar^2+	\frac{1}{2}(A-Y)\right)-16\pi e^{Yr^2}B_g}{3m^2(A+Y) +16\pi e^{Yr^2}B_g}.\nonumber
\end{eqnarray}
The plot of these parameters are shown as follows:
\begin{figure}[H]\centering
	\includegraphics[width=0.45\linewidth]{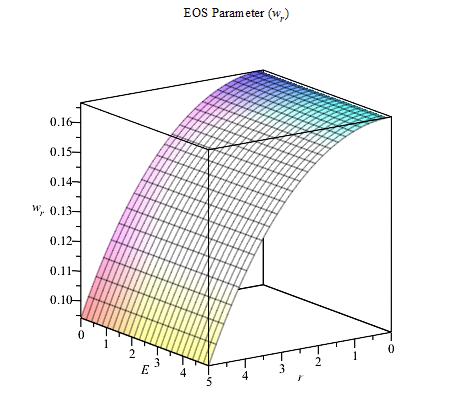}
	\includegraphics[width=0.45\linewidth]{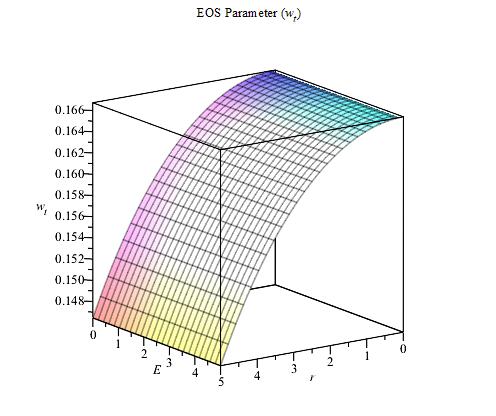}
	\caption{Progression of radial and tangential EOS parameters in RG for $SAX.J1808.4-3658$.}
	\label{f10}
\end{figure}

The EOS radial and transverse parameters, $w_r$ and $w_t$ respectively are shown in FIG. \ref{f10}. The radial parameter is largest in the center and monotonically decreases as we get closer to the star's outer edge. The transverse component grows monotonically from the star's center to its outer edge, with the transverse component being lowest at the center. These two choices both take values in the range of $[0,0.15].$

\begin{figure}[H]
	\centering
	\includegraphics[width=0.35\linewidth]{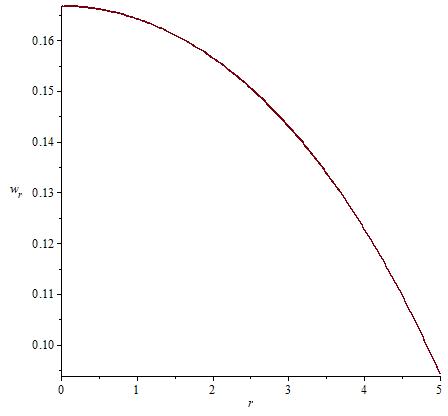}
	\includegraphics[width=0.35\linewidth]{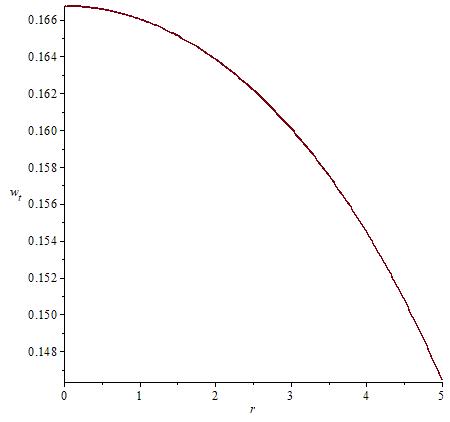}
	\caption{Progression of radial and tangential EOS parameters in GR for $SAX.J1808.4-3658$.}
	\label{f11}
\end{figure}

FIG. \ref{f11} exhibits the two parameters in GR behave similarly.

\subsection{Tolman Openheimer Volkoff Equation}

A compact star's stability is determined using the TOV equation. It presents a thorough examination of the equilibrium condition between gravitational, hydrostatic, and anisotropic forces ($F_g,F_h,F_a$). The vanishing resultant of all system-related forces is required for stable equilibrium. A non-vanishing outcome indicates a system in disequilibrium. 
The following is the general form of the TOV equation:
\begin{equation}
	\frac{dl}{dr}+\frac{2Qr(\rho+p-l)}{r}-2\frac{l}{r}=0.
\end{equation}
Here the gravitational force is $F_g=\frac{dl}{dr},$ the hydrostatic force is $F_h=\frac{2Qr(\rho+p-l)}{r}$ and anisotropic force is $F_a=2\frac{l}{r}$ respectively. For the strange star model, the formulations for these forces are as follows:
\begin{eqnarray}
	\textmd{Gravitational~Force:}F_g&=&\frac{dl}{dr}=-\frac{m^2}{4\pi r^3e^{Yr^2}}(1+Yr^2-e^{Yr^2}-YA(A-Y)r^6\nonumber\\
	&+&(A^2-YA+Y^2)r^4),\\
	\textmd{Hydrostatic~Force:}F_h&=&\frac{Am^2}{2\pi e^{Yr^2}}\left(\frac{e^{Yr^2}-1}{r^2}+A(1+Ar^2-Yr^2)\right),\\
	\textmd{Anisotropic~Force:}F_a&=&\frac{m^2}{4\pi re^{Yr^2}}\left( YAr^2-A^2r^2+Y+\frac{1-e^{Yr^2}}{r^2}\right).
\end{eqnarray}

These forces have the following general relativistic analogue:

\begin{eqnarray}
	\textmd{Gravitational~Force:}F_g|_{E=0}&=&-\frac{1}{4\pi r^3e^{Yr^2}}(1+Yr^2-e^{Yr^2}-YA(A-Y)r^6\nonumber\\
	&+&(A^2-YA+Y^2)r^4),\\
	\textmd{Hydrostatic~Force:}F_h|_{E=0}&=&\frac{A}{2\pi e^{Yr^2}}\left(\frac{e^{Yr^2}-1}{r^2}+A(1+Ar^2-Yr^2)\right)\\
	\textmd{Anisotropic~Force:}F_a|_{E=0}&=&\frac{1}{4\pi re^{Yr^2}}\left( LAr^2-A^2r^2+L+\frac{1-e^{Yr^2}}{r^2}\right)
\end{eqnarray}

The gravitational, hydrostatic, and anisotropic force graphs for $SAX~J1808.4-3658$ are shown below:

\begin{figure}[H]
	\centering
	\includegraphics[width=0.31\linewidth]{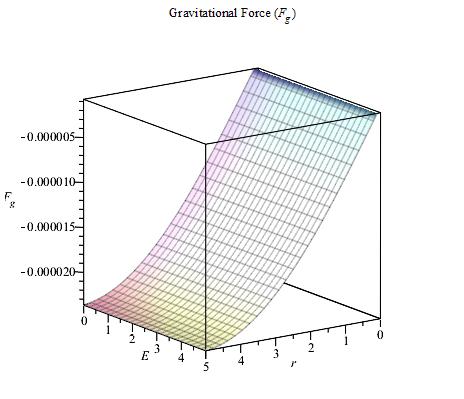}
	\includegraphics[width=0.35\linewidth]{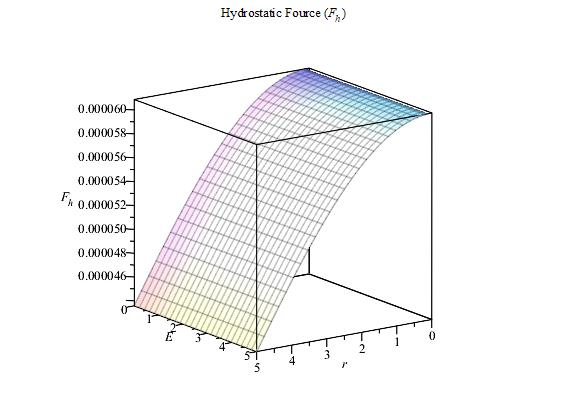}
	\includegraphics[width=0.3\linewidth]{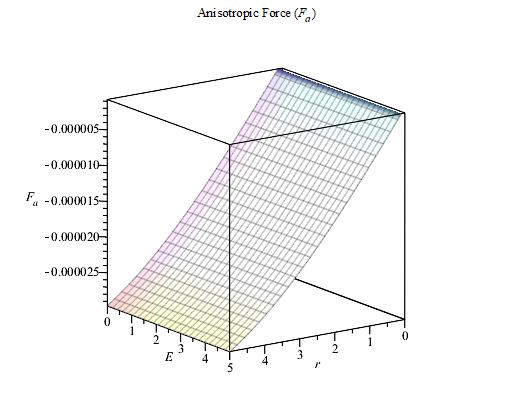}
	\caption{Progression of gravitational, hydrostatic and anisotropic forces in RG for $SAX.J1808.4-3658$.}
	\label{f12}	
\end{figure}

The gravitational, hydrostatic, and anisotropic forces are shown in FIG. \ref{f12}. The graphic unequivocally demonstrates that hydrostatic force is positive whereas gravitational and anisotropic forces are both negative. All three forces are at their greatest in the star's center. They get less intense as we go radially in the direction of the star's periphery. As the energy of the probing particle rises, these forces also do.

\begin{figure}[H]
	\centering
	\includegraphics[width=0.225\linewidth]{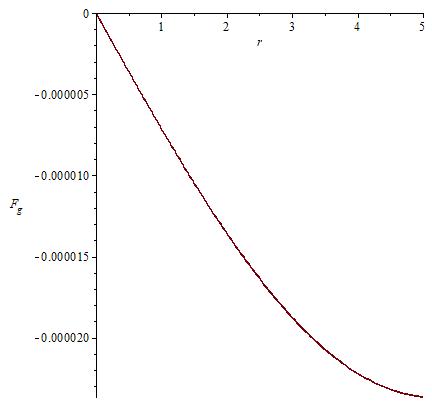}
	\includegraphics[width=0.25\linewidth]{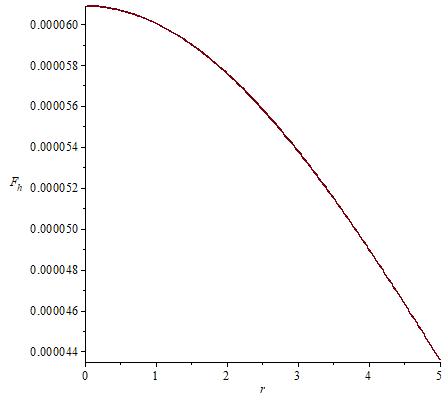}
	\includegraphics[width=0.225\linewidth]{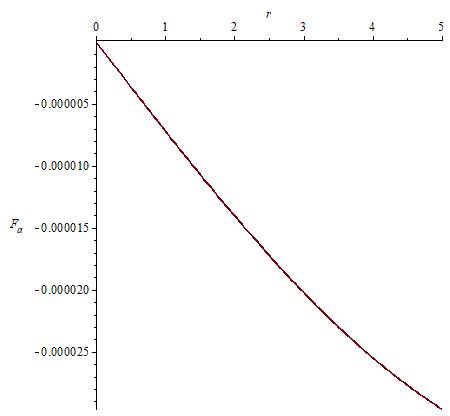}
	\caption{Progression of gravitational, hydrostatic and anisotropic forces in GR for $SAX.J1808.4-3658$.}
	\label{f13}
\end{figure}

FIG. \ref{f13} exhibits a similar behavior of gravitational, hydrostatic and anisotropic forces in GR as RG.

\subsection{Harrera's Cracking Condition}

Harrera's cracking condition explores the stability of strange star configurations. When a stable equilibrium star structure is perturbed, either from incoherent radiation emission or a system anisotropy, the cracking occurs. As a result, it requires a consistent nature for matter distribution. Thus, $v^2_l \leq 1$ is also known as the no-cracking notion. Moreover, literature reveals that $v^2_l$ is also labeled as the squared speed of sound. Here $v^2_l=\frac{dl}{d\rho}$ can be evaluated for the strange star as follows:
\begin{equation}
	v^2_{l}=\frac{dl}{d\rho}=
	\frac{2\left(1+Yr^2-e^{Yr^2}+(A^2-YA+Y^2)r^4-YA(A-Y)r^6\right)}{3Yr^4(Y+A)}
\end{equation}
The general relativity case, i.e., for vanishing energy of probing particles $E,~v^2_{l}$ would be the same as that calculated in RG.

The quantity $v^2_{l}$ in RG and GR is shown as follows:
\begin{figure}[H]
	\centering
	\includegraphics[width=0.4\linewidth]{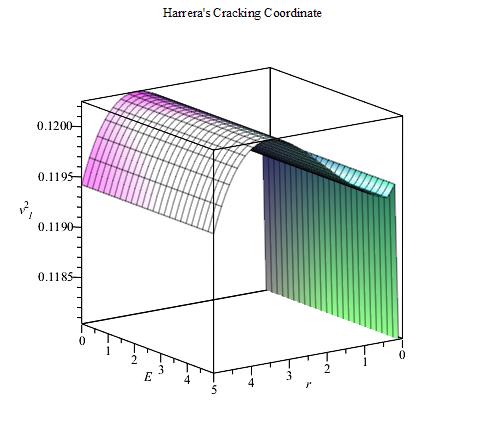}
	\includegraphics[width=0.3\linewidth]{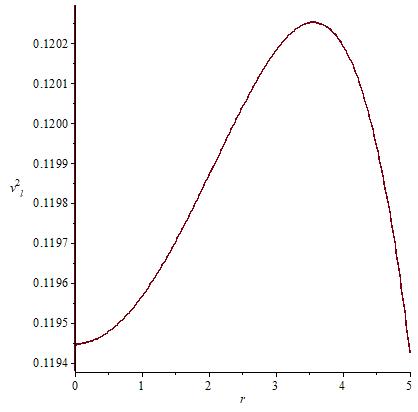}
	\caption{Progression of $v^2_{l}$ in RG and GR for $SAX.J1808.4-3658$.}
	\label{f14}
\end{figure}

The squared speed of sound is greatest near the center of a compact star, as seen in FIG. \ref{f14}. As we move along the radius towards the star's periphery, it gets smaller. The condition does not alter when $E$ changes since the expression is independent of the probing particle's energy. However, across the domain, the squared sound speed lies inside $[0,1]$, supporting Harrera's intriguing theory.

\subsection{Mass Function, Compactness and Redshift}

The strange star model, distinguishing it from a black hole, requires a maximum mass limit for dynamical stability. The following formula can be used to calculate the mass function of a collapsing spherical distribution inside the radius $r$:
\begin{eqnarray}\label{massf}
	M(r)&=&\int_{0}^{r}4\pi r^2 \rho dr,\nonumber\\
	&=&\int_{0}^{r}\left(4\pi r^2 \frac{3m^2}{16\pi e^{Yr^2}}(A+Y)+B_g\right) dr\nonumber\\
	&=&\dfrac{3m^2(A+Y)\left(\sqrt{{\pi}}\,\operatorname{erf}\left(\sqrt{Y}\,r\right)-2\mathrm{e}^{-Yr^2}\sqrt{Y}\,r\right)}{16Y^\frac{3}{2}}+\dfrac{4\pi B_gr^3}{3}
\end{eqnarray}
The mass function in general relativity can be derived for vanishing probing particle energy as:
\begin{equation}\label{massfgr}
	M(r)|_{E=0}\dfrac{3(A+Y)\left(\sqrt{{\pi}}\,\operatorname{erf}\left(\sqrt{Y}\,r\right)-2\mathrm{e}^{-Yr^2}\sqrt{Y}\,r\right)}{16L^\frac{3}{2}}+\dfrac{4\pi B_gr^3}{3}
\end{equation}

In spherical rainbow geometry, the mass function of collapsing string fluid is expressed graphically as:
\begin{figure}[H]
	\centering
	\includegraphics[width=0.5\linewidth]{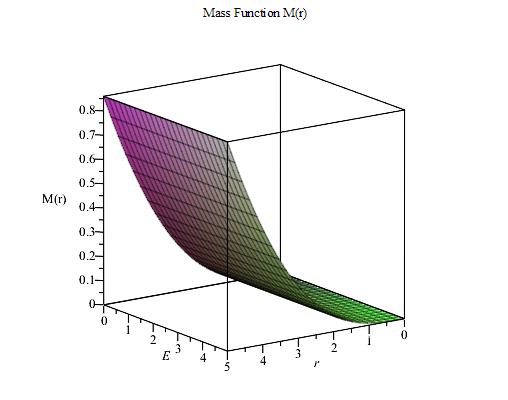}
	\includegraphics[width=0.3\linewidth]{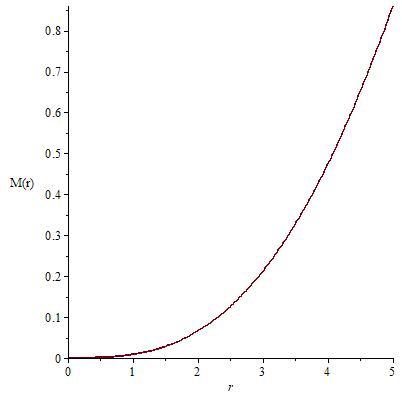}
	\caption{Progression of mass function in RG and GR for $SAX.J1808.4-3658$.}
	\label{f15}
\end{figure}

FIG. \ref{f15} shows that at the strange star's center, the mass function is approximately equal to zero. It grows higher as we approach the star's boundary. The mass function is affected by probing a particle's energy. The mass function grows as the probing particle's energy grows. A low mass in the center shows that the resultant is a strange star rather than a black hole. Furthermore, the function behaves similarly in RG and GR.

The maximal mass of the strange star defines the maximum values of compactness. The compactness factor is defined as the mass function to radius ratio. Compactness of an anisotropic string fluid model is given as:
\begin{eqnarray}\label{CF1}
	u(r)&=&\frac{M}{r} \nonumber \\
	&=&\dfrac{3m^2(A+Y)\left(\sqrt{{\pi}}\,\operatorname{erf}\left(\sqrt{Y}\,r\right)-2\mathrm{e}^{-Yr^2}\sqrt{Y}\,r\right)}{16rY^\frac{3}{2}}+\dfrac{4\pi B_gr^2}{3}
\end{eqnarray}
The general relativistic analogue of the compactness factor can be calculated by reducing the energy of the probing particle to zero, i.e.,
\begin{equation}\label{CF12d}
	u(r)|_{E=0}=\dfrac{3(A+Y)\left(\sqrt{{\pi}}\,\operatorname{erf}\left(\sqrt{Y}\,r\right)-2\mathrm{e}^{-Yr^2}\sqrt{Y}\,r\right)}{16rY^\frac{3}{2}}+\dfrac{4\pi B_gr^2}{3}.
\end{equation}

The compactness factor is represented as:
\begin{figure}[H]
	\centering
	\includegraphics[width=0.5\linewidth]{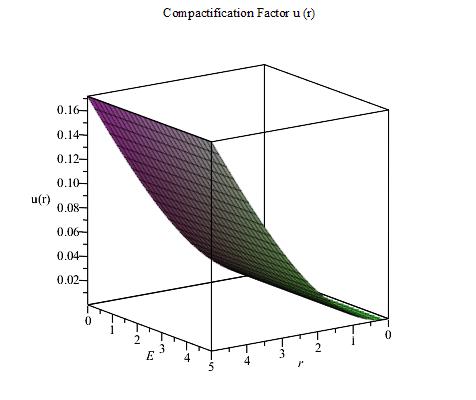}
	\includegraphics[width=0.3\linewidth]{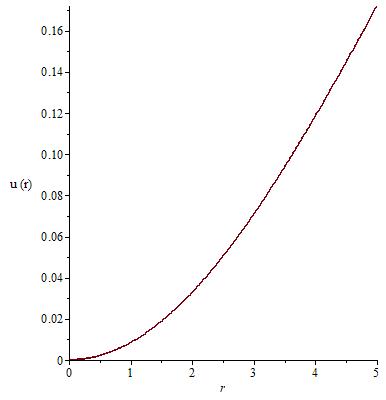}
	\label{f16}
	\caption{Progression of compactness factor in RG and GR for $SAX~J1808.4-3658$.}
\end{figure}

FIG. \ref{f16} exhibits the strange star's compactness factor $SAX~J1808.4-3658.$ The center of the spherical configuration has the lowest compactness factor. This factor increases as $r$ and $E$ increase. The Buchdahl limit, which says that the compactness factor should be less than $\frac{4}{9}\approx 0.444.$, is followed by the compactness factor in RG and GR.

Surface redshift is the term used to describe the shift in electromagnetic radiation caused by a source in the gravitational field, notably near the surface of a peculiar star. The surface redshift can be computed as follows:
\begin{eqnarray}\label{redshift}
	Z_s &=& (1-2u(r))^{-\frac{1}{2}}-1  \nonumber \\
	&=&\left[ 1-\dfrac{3m^2(A+Y)\left(\sqrt{{\pi}}\,\operatorname{erf}\left(\sqrt{Y}\,r\right)-2\mathrm{e}^{-Yr^2}\sqrt{Y}\,r\right)}{8rY^\frac{3}{2}}-\dfrac{8\pi B_gr^2}{3}\right]^{-\frac{1}{2}}-1 \nonumber \\
\end{eqnarray}

The general relativistic analogue of surface redshift can be computed as:
\begin{equation}\label{RSGR}
	Z_s|_{E=0} =\left[ 1-\dfrac{3(A+Y)\left(\sqrt{{\pi}}\,\operatorname{erf}\left(\sqrt{Y}\,r\right)-2\mathrm{e}^{-Yr^2}\sqrt{Y}\,r\right)}{8rL^\frac{3}{2}}-\dfrac{8\pi B_gr^2}{3}\right] ^{-\frac{1}{2}}-1.
\end{equation}

The surface redshift in RG its general relativistic analogue is exhibit as follows:
\begin{figure}[H]
	\centering
	\includegraphics[width=0.5\linewidth]{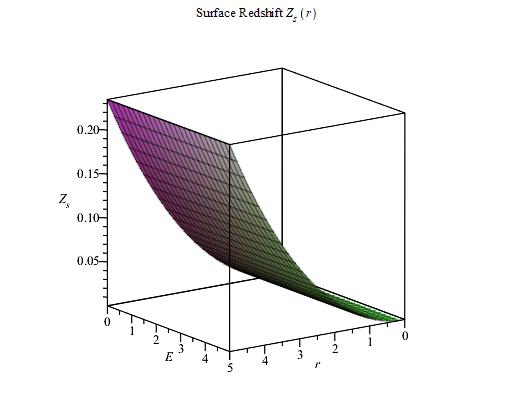}
	\includegraphics[width=0.35\linewidth]{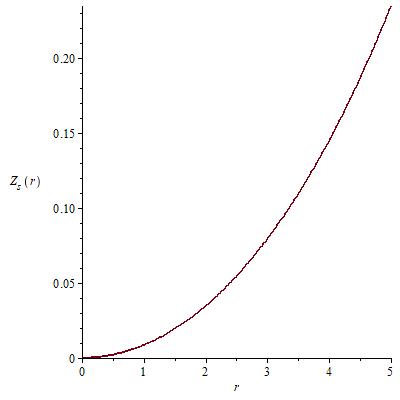}
	\caption{Progression of surface redshift in RG and GR for $SAX.J1808.4-3658$.}
	\label{f17}
\end{figure}
The center of the star has the smallest redshift, as seen in FIG. \ref{f17}. It gets stronger as we go closer to the star's boundary. Furthermore, the redshift grows as $E$ increases. The positive anisotropy factor has no upper limit on the redshift. As a result, the particle with more energy will have a more significant redshift. The redshift behavior is observed to be the same in both GR and RG.

\section{Conclusions}

This research investigates the dynamics of a collapsing charged stringy quark matter under the influence of rainbow gravity. The geometry of the KBR metric is used to understand gravitational collapse and the MIT Bag model is utilized. The real-time data of charged star $SAX~J1808.4-3658$ is used to determine the physical properties like energy conditions, first-order gradients of mass density, pressure and string tension, anisotropy, equation of state parameters, Tolman Oppenheimer Volkoff equation, Harrera's cracking condition, mass compactness, and redshift. The following key findings were obtained:
\begin{itemize}
	\item As one proceeds outward, the mass density and pressure of $SAXJ1808.4-3658$ decrease from their maximum points in the center.
	\item The string tension is negative, reaching a maximum near the center of $SAX~J1808.4-3658$ and lowering as the radius increases.
	\item The squared electric field intensity $\mathbb{E}^2$ is highest at the center of $SAX~J1808.4-3658$ and decreasing as radius increases.
	\item Observers with higher energy experience the dynamical quantities (mass density, pressure, string tension) more intensely compared to those with lower energy.
	\item First-order gradients of dynamical quantities are negative, showing all these functions are decreasing, compatible with a strange star model.
	\item The energy conditions are satisfied for charged stringy quark fluid, supporting its physical presence.
	\item The anisotropy of the fluid increases towards the boundary of the strange star, particularly for observers with higher energy.
	\item The equation of state parameters is positive and larger for particles with higher energy.
	\item The hydrostatic force is positive, but the gravitational and anisotropic forces are both negative, leading to a balanced and stable model of $SAXJ1808.4-3658$.
	\item The squared sound speed is greatest close to the star's centre and falls down towards the edge, with higher energy viewers reporting a quicker speed of sound.
	\item The mass function, compactness, and redshift increase with the increasing radius, with higher values for more energetic observers.
\end{itemize}

General relativistic analogs of all the properties of the RG model for $SAXJ1808.4-3658$ were computed to validate the data, and all properties were determined to be satisfactory. Thus, this study presents a generalized model of charged $SAX~J1808.4-3658$ in RG, incorporating quantum effects through the inclusion of rainbow functions in the expressions of the dynamical quantities of the string fluid. This model supports the evolution and presence of charged strange stars such as $SAX~J1808.4-3658$ in stringy quark matter, probably in the early phases of the Universe.

\section*{Declarations}

\subsection*{Funding}

Not Applicable

\subsection*{Conflicts of Interest/Competing Interests (Include Appropriate Disclosures)}

The authors declare that they have no known competing financial interests or personal relationships that could have appeared to influence the work reported in this paper.
\subsection*{Availability of Data and Material}
Not Applicable
\subsection*{Code Availability}
Not Applicable
\subsection*{Authors' Contributions}
Umber Sheikh: Conceptualization, Design of Analysis, Writing - Reviewing and Editing, Supervision, Software, Validation.\\
Wasib Ali: Data Collection, Methodology, Formal Analysis and Investigations, Writing - Original draft preparation, Investigation, Software, Visualization.
\subsection*{Acknowledgements}
The authors acknowledge University of Education Lahore and HEC Pakistan for the needful support.

\end{document}